\documentclass[final,1p,times]{elsarticle}

\include{elsarticle}

\usepackage{algorithm}
\usepackage{algorithmic}
\usepackage{amssymb}
\usepackage{graphicx}
\usepackage{multirow}
\usepackage{amsmath}
\usepackage{float}

\begin{document}

\begin{frontmatter}

\title{Statistical inference framework for source detection of contagion processes on arbitrary network structures}

\author[naf]{Nino Antulov-Fantulin\corref{cor1}}
\ead{nino.antulov@irb.hr}

\author[al]{Alen Lan\v{c}i\'{c}}
\ead{alen@student.math.hr}

\author[hs,hs2]{Hrvoje \v{S}tefan\v{c}i\'{c}}
\ead{shrvoje@thphys.irb.hr}

\author[ms1,ms2]{Mile \v{S}iki\'{c}}
\ead{mile.sikic@fer.hr}

\author[naf]{Tomislav \v{S}muc}
\ead{tomislav.smuc@irb.hr}

\cortext[cor1]{Corresponding author. Adress: Rudjer Bo\v{s}kovi\'{c} Institute, Bijeni\v{c}ka cesta 54, 10000 Zagreb}

\address[naf]{Computational Biology and Bioinformatics Group, Division of Electronics, \\ Rudjer Bo\v{s}kovi\'{c} Institute, Zagreb, Croatia}
\address[al]{Faculty of Science, Department of Mathematics, University of Zagreb, Zagreb, Croatia}
\address[hs]{Theoretical Physics Division, Rudjer Bo\v{s}kovi\'{c} Institute, Zagreb, Croatia}
\address[hs2]{Catholic University of Croatia, Zagreb, Croatia}
\address[ms1]{Faculty of Electrical Engineering and Computing, Department of Electronic Systems and Information Processing, University of Zagreb, Croatia}
\address[ms2]{Bioinformatics Institute, A*STAR, Singapore, Republic of Singapore}

\begin{abstract}
In this paper we introduce a statistical inference framework for estimating the contagion source from a partially observed contagion spreading process on an arbitrary network structure. The framework is based on a maximum likelihood estimation of a partial epidemic realization and involves large scale simulation of contagion spreading processes from the set of potential source locations. We present a number of different likelihood estimators that are used to determine the conditional probabilities associated to observing partial epidemic realization with particular source location candidates. This statistical inference framework is also applicable for arbitrary compartment contagion spreading processes on networks. We compare estimation accuracy of these approaches in a number of computational experiments performed with the SIR (susceptible-infected-recovered), SI (susceptible-infected) and ISS (ignorant-spreading-stifler) contagion spreading models on synthetic and real-world complex networks. 
\end{abstract}

\end{frontmatter}

The structure of vast majority of biological networks (biochemical, ecological), technological networks (internet, transportation, power grids), social networks and information networks  (citation, WWW) can be represented by complex networks \cite{revNewman}, \cite{revDorogovtsev}, \cite{RevBoccaletti}.
Epidemic or contagion processes are amongst the most prevalent type of dynamic processes of interest characteristic for these real-life complex networks and they include disease epidemics, computer virus spreading, information and rumor propagation \cite{VespignaniNPhys2011}. Different mathematical frameworks have been used to study epidemic spreading. We can divide them into two major categories based upon assumptions they make: the homogeneous mixing framework and the heterogeneous mixing framework. The homogeneous mixing framework assumes that all individuals in a population have an equal probability of contact. This is a traditional mathematical framework \cite{DiffModel1}, \cite{Hethcote} in which differential equations are used to model epidemic dynamics. The heterogeneous mixing framework assumes that properties of contact interactions among individuals are defined via some underlying network structure. The small world network property \cite{SmallWorld} and the scale-free network property \cite{ScaleFree1} \cite{ScaleFree2} have a great impact on the outcome of an epidemic spreading. We can further divide the heterogeneous mixing framework by other assumptions: the bond percolation, the mean-field and the particle network frameworks. The bond percolation approach applies the percolation theory to describe epidemic processes on networks \cite{mollison1977spatial}, \cite{EpiPercolationNet}. The mean-field approach assumes that all nodes having the same degree with respect to an epidemic process are statistically equivalent \cite{Castellano10}, \cite{EpidScaleFree}. The particle network approach assumes that spreading process is characterized by particles which diffuse along edges on a transportation network and each node contains some non-negative integer number of particles (reaction-diffusion processes). 

The main question we address in this work is: Is it possible to detect location of the initial source from partial information on the contagion spread over a network structure ? This research question is useful for many realistic scenarios in which we observe epidemic spread at certain temporal moment and would like to infer the source location (patient-zero). 
Our statistical inference framework is applicable for arbitrary compartment contagion spreading model on arbitrary network structure.
We have based our main case study on the SIR (susceptible-infected-recovered) model but we have demonstrated the applicability of inference framework on other contagion processes like the SI (susceptible-infected) and the ISS (ignorant-spreading-stifler) model. The SIR model \cite{DiffModel1} is an adequate model for many contagious processes like disease modelling, virus propagation \citep{ComputerVirusModels} or rumour propagation \cite{RumorSpreading1}. 
We base our inference study on rather general assumptions which can be relaxed: 
(i) that observed partial epidemic realization is defined by complete knowledge of infected and recovered nodes 
(ii) that probabilities for infection and recovery of the underlying epidemic process are known in advance, as is the time from the start of the epidemic. 
We empirically demonstrate inference performance of the framework on different types of networks and for different contagion properties. 
We also investigate the impact on the performance of the framework in case when the assumptions are relaxed i.e. not complete knowledge on network status and when contagion parameters and time are uncertain. 
Finally, we demonstrate generality of the approach through solving source detection problem for different compartment models (SIR, SI and ISS).

Recently, the problem of estimating the initial source has gained a lot of attention due to its importance and practical aspects.
Under different assumption on network structures or spreading process different source estimators have been developed \cite{Zaman1},\cite{RegularTreeSI_11},\cite{Jaccard_7},\cite{pinto2012},\cite{DMP_0}.
However, we have made a significant contribution in problems of source detection for more general spreading processes on arbitrary network structures.
In this work we cast this problem into a statistical inference framework based on the maximum likelihood estimation of the source of observed epidemic realization. This inference framework relies on a large scale simulation of contagion spreading processes from the set of potential source locations and subgraph similarity measures. 

In section \ref{SIRmodel} we describe the SIR compartment model. Section \ref{StatInference} we describe our statistical inference framework and define different maximum likelihood estimators and subgraph similarity measures used to infer conditional probability of epidemic realizations from particular source locations.
In section \ref{experiments} we describe experiments that demonstrate network, contagion dynamics and noise effects and section \ref{relatedWork} explains the related work.


\section*{Notation}

\begin{tabular}{ | l || l |}
  \hline                        
  Notation & Description   \\
  \hline 
  \hline
$G$ & is a network with a set of nodes $V$ and a set of edges $E$  \\ \hline
$\Theta$ & general variable which identifies source nodes \\ \hline
$\theta$ & specific value for $\Theta$ variable, example: $\Theta = \theta_i$ the source node is the node $i$ \\ \hline
$p$ & probability of infection in one discrete time step \\ \hline
$q$ & probability of recovery in one discrete time step \\ \hline
$n$ & number of simulations for a specific SIR process\\ \hline
$T$ & temporal threshold (random variable or constant)\\ \hline
$\vec{R}$ & epidemic random vector $\vec{R}= (R(1),R(2),..,R(k))$\\ \hline
$R(i)$ & Bernoulli indicator random variable for node $i$\\ \hline
$\vec{r}$ & epidemic realization, example $\vec{r}_1= (1,0,0,1,...,1)$\\ \hline
$\vec{r_*}$ & observed epidemic realization\\ \hline
$\vec{r}(i)$ & i-th component of the realization vector $\vec{r}$, 
example $\vec{r}= (1,0,1,1)$,  $\vec{r}(2) = 0$\\ \hline
$\vec{R}_{\theta}$ & random vector for realizations from node $\theta$\\ \hline
$\vec{R}_{\theta,i}$ & i-th sample realization vector from random vector $\vec{R}_{\theta}$\\ \hline
$S$ & set of potential sources\\ \hline
$\varphi(\vec{r}_1, \vec{r}_2)$ & similarity measure between two realizations $\vec{r}_1, \vec{r}_2$\\ \hline
$\varphi(\vec{r}_*, \vec{R}_\theta)$ & random variable which measures the similarity between \\ 
& realization $\vec{r}_*$ and realizations from random vector $\vec{R}_\theta$\\ \hline
$\psi_{\oplus}(m_1,m_2)$ & a bitwise XNOR function\\ \hline
$\psi_{\vee}(m_1,m_2)$ & a bitwise OR function\\ \hline
$\psi_{\wedge}(m_1,m_2)$ & a bitwise AND function\\ \hline
$\varphi_x(\vec{r_1},\vec{r_2})$ & the similarity calculated with $\overline{XNOR}(\vec{r_1},\vec{r_2})$ function \\ \hline
$\varphi_J(\vec{r_1},\vec{r_2})$ & the similarity calculated with $Jaccard(\vec{r_1},\vec{r_2})$ function\\ \hline
$\delta(x)$ & the Dirac delta function \\ \hline
\hline  
\end{tabular}

\section{SIR compartment model}
\label{SIRmodel}
We define the contact-network as an undirected and non-weighted graph $G(V,E)$ ($V$-set of nodes or vertices, $E$-set of links). A link $(u,v)$ exists only if two nodes $u$ and $v$ are in contact during the epidemic time. We also assume that the contact-network during the epidemic process is a static one. 
To simulate epidemic propagation through a contact-network, we use the standard stochastic SIR model. In this model each node at some time can be in one of the following states: susceptible (S), infected (I) and recovered (R). The spreading process is simulated using discrete time step model.

The SIR epidemic process is a stochastic process, which is simulated with $n$ mutually independent simulation steps on the contact network $G$.
At the beginning of each epidemic simulation all nodes from graph $G$ are in the susceptible state except set of nodes which are initially infected. We assume in our treatment that epidemic parameters $p$ and $q$ are predefined, constant and known beforehand. The epidemic parameter $p$ is the probability that an infected node $u$ infects an adjacent susceptible node $v$ in one discrete time step. The epidemic parameter $q$ is the probability that an infected node recovers in one discrete time step. Set of initially infected nodes is denoted with the letter $\Theta$.  At the end of one full epidemic simulation, all nodes can be in one of two following states: susceptible or recovered. In our treatment however, we will limit epidemic spreading to a predefined number of discrete time steps, which basically means that we will deal with partially realized epidemic spreads and that this number of steps is also known parameter in the inference procedure for the source location estimation.

\section{Statistical inference on epidemic propagation realizations}
\label{StatInference}
In this section we formulate the problem of the source localization in the network and develop related statistical inference framework.

\subsection*{\textbf{Epidemic source location problem}}
Let us define the random vector $\vec{R}=(R(1),R(2),...,R(N))$, that indicates which nodes got infected prior up to some predefined temporal threshold $T$ (random variable or constant). 
The random variable $R(i)$ is a Bernoulli random variable, which assigns the value $1$ if node $i$ got infected before time $T$ from the start of the epidemic process and the value $0$ otherwise. \\

Let us assume that we have observed one spatio-temporal epidemic propagation realization $\vec{r}$ of SIR process defined by $(p,q)$ and $T$, and we want to infer which nodes from the set $S$ are the most likely source of realization $\vec{r}$ for the SIR process $(p,q)$ and $T$.  $S=\left\lbrace \theta_1, \theta_2, ..., \theta_N \right\rbrace$ is the finite set of possible source nodes that is defined by observed infected or infected and recovered set of nodes prior to moment $T$ in the network.


In order to find a node or a small subset of infected nodes that have highest likelihood for being the source of the epidemic spread, we pose the following maximum likelihood problem.
\[
\hat{\Theta} = \arg\max_{\Theta \in S} P(\Theta|\vec{R}=\vec{r}),
\]
where $\Theta \in S$ is a set of all possible sources of epidemic. \\
By applying Bayes theorem, we get the following expression:
\[
\hat{\Theta} = \arg\max_{\Theta \in S} \frac{P(\vec{R}=\vec{r}|\Theta=\theta)P(\Theta=\theta)}{\sum_{\theta_k} P(\vec{R}=\vec{r}|\Theta=\theta_k)P(\Theta=\theta_k)}. 
\]
If all $\Theta$ (apriori) are equally likely, this is equivalent to:
\[
\hat{\Theta} = \arg\max_{\Theta \in S} P(\vec{R}=\vec{r}|\Theta=\theta).
\]

Thus, the core of source location estimation problem is the determination of the likelihood 
of the observed epidemic realization being initiated at the source location $\Theta$. We now proceed with description of the algorithms for determining the maximum likelihood for the observed epidemic realization.

\subsection{The Maximum Likelihood source estimator}
First, we give a pseudo-code (Algorithm 1) for the original problem of the maximum likelihood source estimation, where source can be any node from set $S$. In principle, this treatment can be extended to problem of multiple sources determination, but the necessary extensions are out of the scope of this work. Note that among algorithm parameters $(G,p,q,\vec{r}_{*},T,S,n)$ the parameter $n$ represents number of random simulations from a
single candidate for the epidemic source node. In our framework, the number of random simulations $n$ is very important from the perspective of the accuracy/stability of results and it is also a major determinant of the running time of the estimation procedure.

\begin{algorithm}[H]
\label{alg:ML-alg}
\caption{The Maximum Likelihood source estimator algorithm: $(G,p,q,\vec{r}_{*},T,S,n)$}
\begin{algorithmic}
\STATE \textbf{Input:} Network structure $G$, SIR process parameters $(p,q)$, $S=\left\lbrace \theta_1, \theta_2, ..., \theta_N \right\rbrace$ a set of possible sources $\theta_i$, observed realization $\vec{r}_{*}$ ending at some temporal threshold $T$ , $n$ a number of simulations
\FOR{each $\theta_j \in S$ (apriori set of possible sources of epidemic)}
\STATE Call likelihood estimation function $(G,p,q,\vec{r}_{*},T,n)$ 
\STATE Save $\hat{P}(\vec{R}= \vec{r}_{*}|\Theta=\theta_j)$
\ENDFOR
\STATE \textbf{Output 1:} $\theta_k$ with maximum likelihood $\hat{P}(\vec{R}= \vec{r}_{*}|\Theta=\theta_k)$ 
\STATE \textbf{Output 2:} Ranked sources in $S=\left\lbrace \theta_1, \theta_2, ..., \theta_N \right\rbrace$ according to likelihoods $\hat{P}(\vec{R}= \vec{r}_{*}|\Theta=\theta_k)$
\end{algorithmic}
\end{algorithm}

It is obvious that the Algorithm \ref{alg:ML-alg} is just a wrapper code that calls likelihood estimation function for each potential source of epidemics. We now proceed with the description of different algorithms for calculating the likelihood $P(\vec{R}=\vec{r}|\Theta=\theta)$. \\

\subsection{Realization similarity matching}
Let us define the function $\varphi(\vec{r_1},\vec{r_2})$, which measures the similarity  between two epidemic realizations or subgraphs of the underlying network: $\vec{r_1}$ and $\vec{r_2}$.

We first define new random variable $\varphi(\vec{r_*}, \vec{r_{\theta}} )$, which measures the $\varphi$ similarity between the fixed realization $\vec{r_*}$ and random realization that comes from $SIR$ process with the source $\theta$. We can calculate the unbiased estimator of the following cumulative distribution function as the empirical distribution function:
\[
\hat{F}(x) = \hat{P}( \varphi(\vec{r_*}, \vec{R_{\theta}} ) \leq x ) = \frac{\sum_{i=1}^n \textbf{1}_{[ 0,x\rangle} \left( \varphi(\vec{r_*}, \vec{R}_{\theta,i}) \right) } {n},
\]
where $\textbf{1}_{[ 0,x\rangle}$ is a characteristic function defined as:
\[
\textbf{1}_{[ 0,x\rangle}(y) = \left\{ 
  \begin{array}{l l}
   1 & \quad \text{: $y \in [0,x\rangle$},\\
   0 & \quad :else.\\
  \end{array} \right.
\]
Then, its probability density function is calculated like this:
\[
PDF(x) = \frac{d}{dx} \hat{F}(x) = \frac{1}{n} \sum_{i=1}^n \delta \left( x- \varphi(\vec{r_*}, \vec{R}_{\theta,i}) \right),
\]
where $\delta(x)$ is the Dirac delta function.  \\
Central limit theorem states that pointwise, $\hat{F}(x)$ has asymptotically normal distribution. The rate at which this convergence happens is bounded 
by Berry--Esseen theorem. This implies that the rate of convergence is bounded by $O(1 \/ /\sqrt{n})$, where n is the number of random simulations.

Next, we define two measures ($XNOR$ and Jaccard) that are used to determine the similarity $\varphi$. 
The first one is a binary NOT XOR function or $XNOR(\vec{r_1},\vec{r_2})$ counts the number of corresponding non-infected and infected nodes in realizations $\vec{r_1}$ and $\vec{r_2}$:
\[
XNOR(\vec{r_1},\vec{r_2}) = \sum_{k \in V} \psi_{\oplus} ( \vec{r_1}(k), \vec{r_2}(k) ),
\]
,where $\psi_{\oplus}(m_1,m_2)$ function is defined as:
\[
\psi_{\oplus}(m_1,m_2) = \left\{ 
  \begin{array}{l l}
   1 & \quad \text{: ($m_1$ = $1$ and $m_2$ = $1$) or ($m_1 = 0$ and $m_2 = 0$) },\\
   0 & \quad \text{: else}.\\
  \end{array} \right.
\]
In other words, $\psi(m_1,m_2)$ is equal to one only if two nodes were infected or they did not get infected prior to temporal threshold $T$. 
We also define function: $\overline{XNOR}(\vec{r_1},\vec{r_2})$, which is normalized $XNOR$ function over total number of nodes: $XNOR(\vec{r_1},\vec{r_2})*N^{-1}$. 

The second similarity measure is a well known Jaccard measure, which in our case counts the number of corresponding infected nodes in $\vec{r_1}$ and in $\vec{r_2}$ normalized by the number of corresponding infected nodes in $\vec{r_1}$ or in $\vec{r_2}$.
\[
Jaccard(\vec{r_1},\vec{r_2}) = \frac{| \vec{r_1} \wedge \vec{r_2}|}{| \vec{r_1} \vee \vec{r_2} |} = 
\frac{\sum_{k \in V} \psi_{\wedge} ( \vec{r_1}(k), \vec{r_2}(k) ) }{\sum_{k \in V} \psi_{\vee} ( \vec{r_1}(k), \vec{r_2}(k) ) },
\]
where $\psi_{\wedge}(m_1,m_2)$ and $\psi_{\vee}(m_1,m_2)$  functions are defined as:
\[
\psi_{\wedge}(m_1,m_2) = \left\{ 
  \begin{array}{l l}
   1 & \quad \text{: ($m_1$ = $1$ and $m_2$ = $1$)},\\
   0 & \quad \text{: else}\\
  \end{array} \right.
\]
and where $\psi_{\vee}(m_1,m_2)$  function is defined as:
\[
\psi_{\vee}(m_1,m_2) = \left\{ 
  \begin{array}{l l}
   1 & \quad \text{: ($m_1$ = $1$ or $m_2$ = $1$) },\\
   0 & \quad \text{: else}.\\
  \end{array} \right.
\]

In the following text the $\varphi_x(\vec{r_1},\vec{r_2})$ will denote the similarity calculated with $\overline{XNOR}(\vec{r_1},\vec{r_2})$ function and 
$\varphi_J(\vec{r_1},\vec{r_2})$ will denote the similarity calculated with $Jaccard(\vec{r_1},\vec{r_2})$ function. In order to speed the similarity matching between realizations, we use the bitwise operations (XOR, NOT, AND) and bit count with Biran-Kernignan method.

\subsection{Likelihood estimation functions}

In this section we define three variants of likelihood estimation functions: AUCDF, AvgTopK, and Naive Bayes. First two functions, AUCDF and AvgTopK can use any of the similarity measures defined above, while Naive Bayes produces likelihood based on its own similarity measure.


As a first likelihood estimation function we define AUCDF (Area Under Cumulative Distribution Function)  (see Algorithm \ref{alg:EstFun1}), which can use any of the similarity measures defined above. 

\begin{algorithm}[H]
\caption{AUCDF estimation function $(G,p,q,\vec{r}_{*},T, \theta,n)$}  
\label{alg:EstFun1}
\begin{algorithmic}
\STATE \textbf{Input:} $G$ - network structure , $(p,q)$ - SIR process parameters , $\vec{r}_{*}$ - observed realization prior to some temporal threshold $T$, $\theta$ - source for which likelihood is calculated, $n$ a number of simulations
\FOR{$i$ = 1 to $n$ (number of simulations)} 
\STATE - Run SIR simulation $(p,q)$ with $\Theta=\theta$ and obtain epidemic realization $\vec{R}_{\theta,i}$, ending at the temporal threshold $T$; 
\STATE - Calculate and save $\varphi(\vec{r}_{*},\vec{R}_{\theta,i})$ ;
\ENDFOR
\STATE - Calculate empirical distribution function:
\[
\hat{P}( \varphi(\vec{r_*}, \vec{R_{\theta}} ) \leq x ) = \frac{\sum_{i=1}^n \textbf{1}_{[ 0,x\rangle} ( \varphi(\vec{r_*}, \vec{R}_{\theta,i}) ) } {n}
\]
\STATE - Estimate likelihood using the area under the empirical cumulative distribution:
\[
AUCDF_{\theta} = \int_0^1 \! \hat{P}( \varphi(\vec{r_*}, \vec{R_{\theta}} ) \leq x ) \mathrm{d} x
\]
\STATE \textbf{Output:} $\hat{P}(\vec{R}=\vec{r}_{*}|\Theta=\theta) = 1-AUCDF_\theta$ likelihood for $\theta$;
\end{algorithmic}
\end{algorithm}

Different sources $\theta$ produce different empirical cumulative distributions of similarities to $\vec{r_*}$. If we compare two empirical distribution functions $CDF_{1}$ and $CDF_{2}$ from two different sources $\theta_1$ and $\theta_2$ and if the $AUCDF_1 < AUCDF_2$ then sample of realizations from $\theta_1$ source are more similar to fixed realization $\vec{r_*}$ than the sample realizations from $\theta_2$ source. This is the primary reason, why we use value $1-AUCDF$ to estimate source likelihood  $\hat{P}(\vec{R}=\vec{r}_{*}|\Theta=\theta)$.


Algorithm AvgTopK represents a variant of the previous estimation function, which uses only $k$ highest values from the tail of the probability density function of the random variable $\varphi(\vec{r_*}, \vec{r_{\theta}})$:

\[
PDF(x) = \frac{d}{dx} \hat{F}(x) = \frac{1}{n} \sum_{i=1}^n \delta \left( x- \varphi(\vec{r_*}, \vec{R}_{\theta,i}) \right).
\]

\begin{algorithm}[H]
\caption{AvgTopK likelihood estimation function $(G,p,q,\vec{r}_{*},T, \theta,n)$}  
\label{alg:EstFun2}
\begin{algorithmic}
\STATE \textbf{Input:} $G$ - network structure , $(p,q)$ - SIR process parameters , $\vec{r}_{*}$ - observed realization prior to some temporal threshold $T$, $\theta$ - source for which likelihood is calculated, $n$ a number of simulations
\FOR{$i$ = 1 to $n$ (number of simulations)} 
\STATE - Run SIR simulation $(p,q)$ with $\Theta=\theta$ and obtain epidemic realization $\vec{R}_{\theta,i}$, ending at the temporal threshold $T$; 
\STATE - Calculate and save $\varphi(\vec{r}_{*}, \vec{R}_{\theta,i})$ ;
\ENDFOR
\STATE - Sort the scores $\left\{ \varphi(\vec{r}_{*}, \vec{R}_{\theta,i}) \right\}$ in descending order;
\STATE - Average top $k$ highest scores:
\[
\hat{P}(\vec{R}=\vec{r}_{*}|\Theta=\theta) = \frac{1}{k} \sum_{i=1}^k \left\{ \varphi(\vec{r}_{*}, \vec{R}_{\theta,i})) \right\}_{sorted}
\]
\STATE \textbf{Output:} $\hat{P}(\vec{R}=\vec{r}_{*}|\Theta=\theta)$ likelihood for $\theta$;
\end{algorithmic}
\end{algorithm}

In each simulation we calculate how similar $\vec{R}_{\theta,i}$ realization to observed $\vec{r}_{*}$ realization is by using $\varphi$ function. 
Estimate $\hat{P}(\vec{R}=\vec{r}_{*}|\Theta=\theta)$ is the average score over top $k$ highest similarities $\varphi(\vec{r}_{*}, \vec{R}_{\theta,i})$ in $n$ simulations (tail of pdf). 

Finally, we propose the third likelihood estimation function which is based on node probabilities for being infected from a particular source node. Main assumption of this approach is independence between nodes with respect to epidemic spreading.

The conditional probability that the node $k$ in realization $\vec{r}_*$ is infected from source $\theta$ is:
\[
\hat{P}(\vec{r}_*(k) = 1 | \Theta=\theta) = \frac{m_k+\epsilon}{n+\epsilon}, \forall k \in G,
\]
where $m_k$ is the number of times that node $k$ got infected from the total of $n$ simulations $SIR(p,q)$ from source node $\theta$ and $\epsilon$ is a smoothing factor. Smoothing factor $\epsilon$ is necessary to mitigate the problem of zero values, stemming from the finite number of simulations used to calculate $\hat{P}(\vec{r}_*(k) = 1 | \Theta=\theta)$.

Then we define the estimator for the likelihood of observing realization $\vec{r_*}$ from source node $\theta$ as:
\[
\hat{P}(\vec{R}=\vec{r_*}|\Theta=\theta) = \prod_{\{ k : \vec{r_*}(k) = 1 \}} \hat{P}(\vec{r}_*(k) = 1  | \Theta=\theta) \prod_{\{ j : \vec{r_*}(j) = 0 \}} (1-\hat{P}(\vec{r}_*(j)| \Theta=\theta)).
\]

This equation uses the probability estimates that nodes $\{ k : \vec{r_*}(k) = 1 \}$ from realization $\vec{r_*}$ got infected and the probability estimates that nodes $\{ j : \vec{r_*}(j) = 0 \}$ from realization $\vec{r_*}$ did not get infected from source node $\theta$.

In mathematical sense, probability of finding an infected node $k$ at time $t$ is dependent on other infected nodes prior to time $t$. Nevertheless, we use the same assumption of independence to estimate the rank of potential sources. There is obvious resemblance between this approach and the well known studied probabilistic classifier - Naive Bayes. Although Naive Bayes uses a strong assumption of independence, it has been shown that in practice its performance is comparable to more complex probabilistic classifiers \cite{NaiveBayes1}. \\


In order to have more stable numerical likelihood estimations, we used the log likelihood variant for estimating $\hat{P}(\vec{R}=\vec{r}_*|\Theta=\theta))$ (see Algorithm \ref{alg:EstFun2}).

\begin{algorithm}[H]
\caption{Naive Bayes likelihood estimation function $(G,p,q,\vec{r}_{*},T,\theta, n)$ }  
\label{alg:EstFun3}
\begin{algorithmic}
\STATE \textbf{Input:} $G$ - network structure , $(p,q)$ - SIR process parameters , $\vec{r}_{*}$ - observed realization prior to some temporal threshold $T$, $\theta$ - initial source for which likelihood is calculated
\STATE - $m_k = 0:  \forall k \in V$ from $G$;
\FOR{$i$ = 1 to $n$ (number of simulations)} 
\STATE - Run SIR simulation $(p,q)$ with $\Theta=\theta$ and obtain realization $\vec{R}_{\theta,i}$ prior to the temporal threshold $T$; 
\STATE - Update: $m_k$ = $m_k + 1;$ $\forall k$ which were infected in $\vec{R}_{\theta,i}$;
\ENDFOR
\STATE - Calculate: 
\[
\hat{P}(\vec{r}_*(k) = 1 | \Theta=\theta) = \frac{m_k+\epsilon}{n+\epsilon}, \forall k \in G
\]
\STATE - Calculate log likelihood: $log(\hat{P}(\vec{R}=\vec{r}_*|\Theta=\theta)) =$ 
\[
= \sum_{\{ k : \vec{r_*}(k) = 1 \}} log(\hat{P}(\vec{r}_*(k) = 1  | \Theta=\theta)) + \sum_{\{ j : \vec{r_*}(j) = 0 \}} log(1-\hat{P}(\vec{r}_*(j)| \Theta=\theta));
\]
\STATE \textbf{Output:}  $log(\hat{P}(\vec{R}=\vec{r}_*|\Theta=\theta))$ likelihood for $\theta$;
\end{algorithmic}
\end{algorithm}

\section{Epidemic source location experiments}
\label{experiments}
In this section, we describe the experiments along with the obtained results performed on different networks and in different epidemic settings. The experiments were designed to illustrate the overall predictability properties of the source detection problem with the introduced inference framework and compare the performances of individual algorithms. \\

\begin{figure}[H]
\caption{ One epidemic realization of SIR process $(p = 0.3, q = 0.7)$ on a synthetic grid, where the color gradient from blue to red represents estimated source likelihood (blue - lower and red - higher ). Node with the letter "A" represents true source of epidemic realization and the node with the letter "B" represents the Maximum Likelihood source estimate by the "Naive Bayes" likelihood estimation function}

\begin{center}
\includegraphics[scale=0.3]{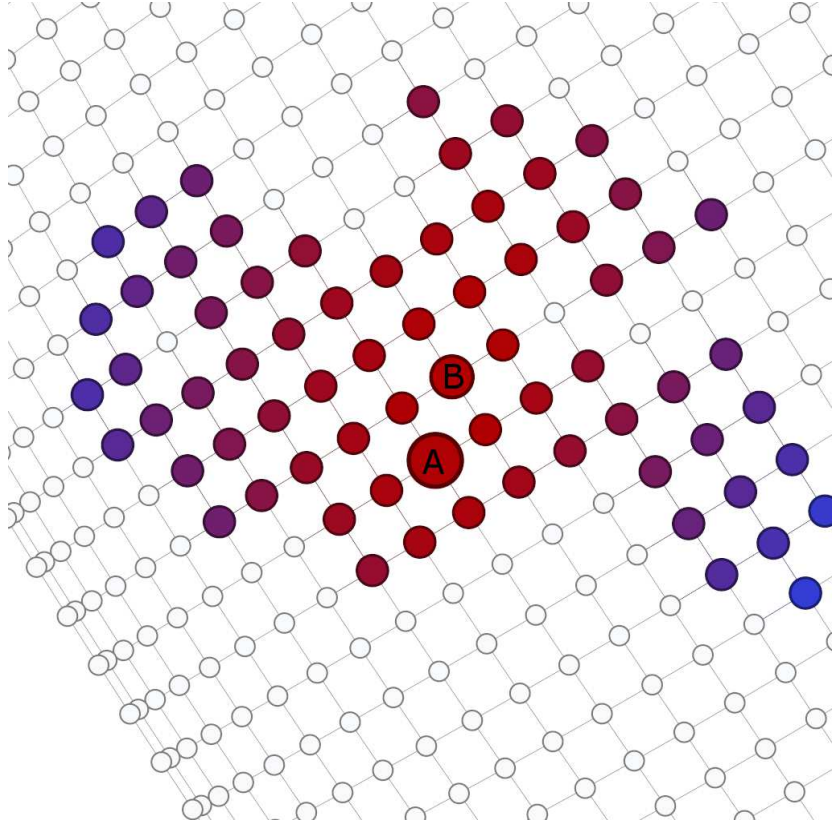} 
\end{center}
\label{fig:realizationGrid}
\end{figure}

We test the performance of source likelihood estimation algorithms on single source epidemic detection problems. In our experiments we observe
one spatio-temporal epidemic propagation realization $\vec{r_*}$ and we want to infer the potential source of realization from the set $S$. In Figure \ref{fig:realizationGrid}, we illustrate one epidemic realization on a synthetic grid, where the color gradient from blue to red represents estimated source likelihood (blue - lower and red - higher ). We have used a Naive SIR algorithm implementation \cite{FastSIR} as efficient SIR process simulation on network structures.

Due to the strong stochastic nature of epidemic process, frequency of correct estimations of the source location in the
network is not the best measure to test the predictability of algorithms. 
The topological distance of maximum likelihood node from true source can be a misleading low even for random estimations on networks with low average shortest path.
Therefore, we measure the rank of true source in the output list of potential sources from set $S$ in experiments on different network structures. The overall testing procedure is given in the following pseudo-code.

\begin{algorithm}
\label{alg:SourceLocationExperimentCode}
\caption{Source location experiments}
\begin{algorithmic}
\FOR{experiment = 1 to total number of experiments}
\STATE - Sample random initial source  $\theta_*$ from network $G$
\STATE - Obtain realization $\vec{r}_*$ from SIR process $SIR(p,q,\Theta = \theta_* ,T)$ that has at least $0.01$ infected nodes in total network
\STATE - Create set $S$ as the set of all nodes which were infected in realization $\vec{r}_*$
\STATE - Call the Maximum Likelihood source estimator algorithm $(G,p,q,\vec{r}_{*},T,S)$
\STATE - Measure the rank of true source on ranked likelihood list of $S$
\ENDFOR
\end{algorithmic}
\end{algorithm}

Let us assume that in some source location detection experiment we get realization $\vec{r}_*$ that has $k$ infected nodes. We rank the nodes $\theta_i$ in a list of $k$ potential nodes according to the likelihood $\hat{P}(\vec{R}=\vec{r}_*|\Theta=\theta_i)$. We express the rank of real source as a relative source rank, i.e. the rank of the true source node normalized to the list size (for example, if the rank of the true source node is at the position $10$ in the list of $100$ potential sources, then the relative source rank is $0.1$). 
For the performed batch of experiments, we calculate cumulative source rank probability distribution, which tells us the probability that the relative rank of the source node is lower or equal to some specified value. 
By its nature cumulative source rank is very similar to the well known receiver operating characteristic (ROC), a measure frequently used in signal detection and machine learning for measuring the performance of classifier systems. Ideal estimator or classifier would have area under the cumulative source rank equal to one, exactly as in the case of ROC measure (AUC measure represents the area under the ROC curve).
One can argue that other measures might have been appropriate as well, for example the distance of the maximum likelihood node to the true source node in a network. We opted for cumulative source rank, because it is a more versatile measure, due to its invariance to network size size and structure (e.g. for networks with different average shortest paths one would get grossly different results).

\begin{figure}[]
\caption{ Visualization of regular grid of size $N = 30x30$}
\begin{center}
\includegraphics[scale=0.4]{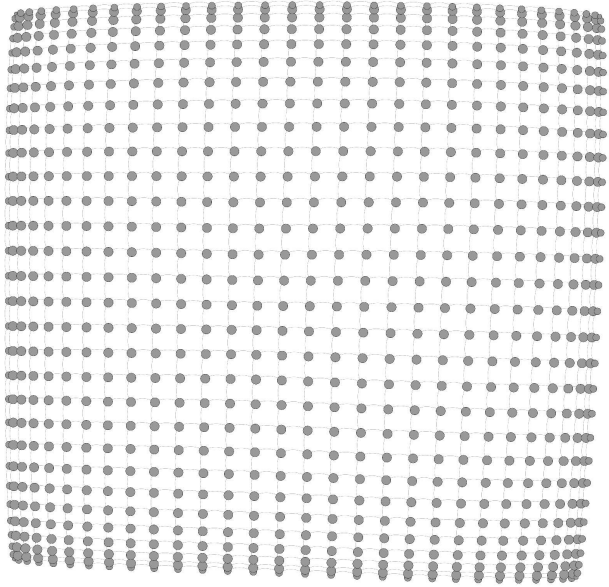} 
\end{center}
\label{fig:RegularGridVis}
\end{figure}

\begin{figure}[]
\caption{ Visualization of Albert-Barabasi network (part A) of size $N = 5000$, with $m_0=5$ initial full connected core, and $m=1$ added edges in preferential attachment. In part B: the visualization of power-grid network (Western States Power Grid of the United States \cite{SmallWorld}) of size $N = 4941$.}
\begin{center}
\includegraphics[scale=0.35]{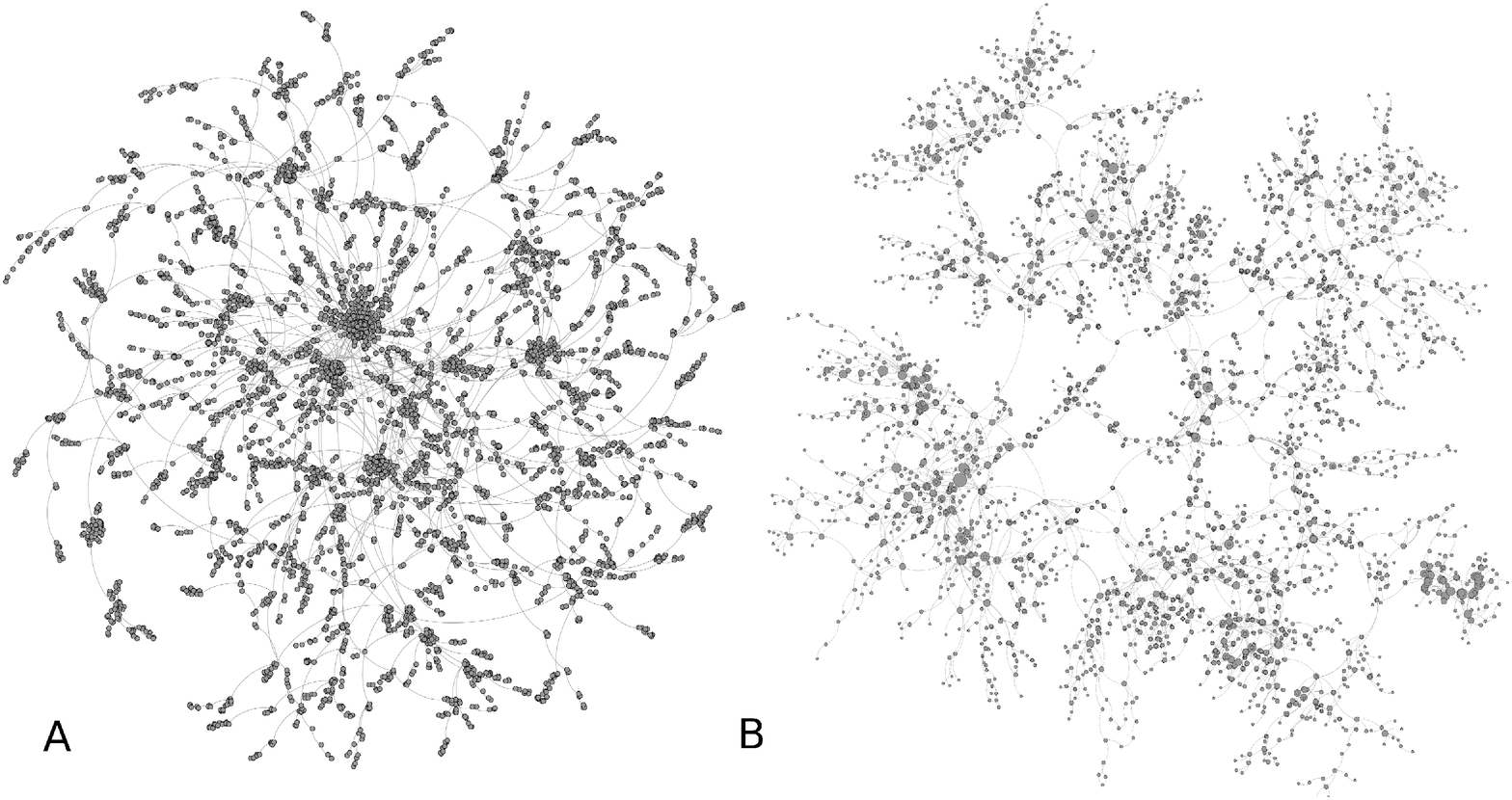} 
\end{center}
\label{fig:AB_Power_Viz}
\end{figure}

\begin{figure}[H]
\caption{ Classes of networks are generated according to the Watts-Strogatz small-world $\beta$ model (size: $N = 5000$) from the regular lattice ($\beta = 0$ and 10 local edges) to random networks ($\beta = 1$) with small-world networks in the middle. Visualization is done on smaller networks (size: 50) from regular lattice to random networks (part C: $\beta = 1$) with Small-world networks in the middle (e.g. part B: $\beta = 0.1$).}
\begin{center}
\includegraphics[scale=0.7]{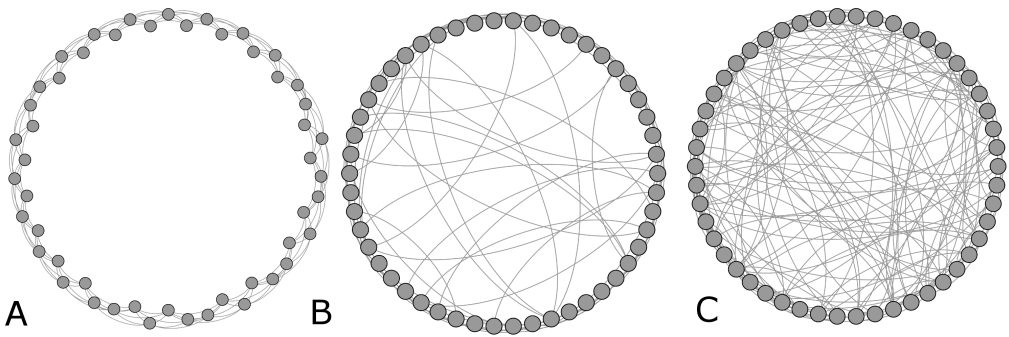} 
\end{center}
\label{fig:SW_vis}
\end{figure}

The influence of network structure on source localization performance has been tested on the following classes of networks: regular grid (figure \ref{fig:RegularGridVis})  and lattice (figure \ref{fig:SW_vis} part A), Small-World networks (figure \ref{fig:SW_vis} part B), Erd\"{o}s-R\'{e}nyi networks (figure \ref{fig:SW_vis} part C), Albert-Barabasi network (figure \ref{fig:AB_Power_Viz} part A) and Western States Power Grid of the United States \cite{SmallWorld} (figure \ref{fig:AB_Power_Viz} part B).

In order to measure the performance of source localization we have done the following experiments:
\begin{itemize}
\item Comparison of different estimators: we compare performance of different algorithms for different epidemic conditions,
\item Network structure experiments: this set of experiments illustrates the effects of network structure on the prediction performance over diverse network topologies,
\item Process dynamics experiments: here we observe the effects of different process parameters like $(p,q,T)$ on source localization performance and
\item Uncertainty experiments: Performance degradation associated with uncertain epidemic parameters or incomplete knowledge about epidemic realization.
\end{itemize}

\subsection*{Comparison of different estimators}

\begin{figure}[H]
\caption{Cumulative probability distribution of source relative rank based on 500 experiments with random initial source on synthetic grid $N = 30x30$ for $p=0.3$, $q=0.7$, $T=10$ with different likelihood estimation functions.}
\begin{center}
\includegraphics[scale=0.45]{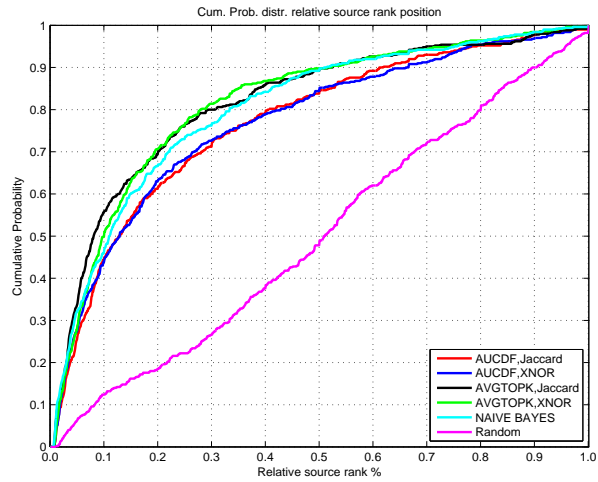} 
\end{center}
\label{fig:SourceRankGrid}
\end{figure}

\begin{figure}[H]
\caption{Cumulative probability distribution of source relative rank for 500 experiments with random initial source on power grid network of size $N = 4941$ for $p=0.7$, $q=0.6$, $T=7$ and different likelihood estimation functions.}
\begin{center}
\includegraphics[scale=0.45]{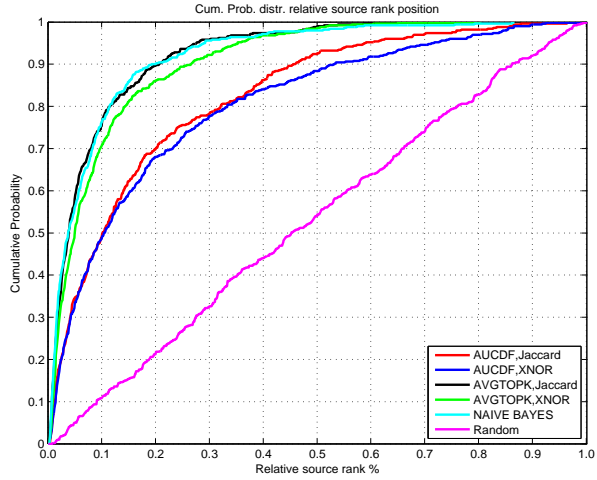} 
\end{center}
\label{fig:SourceRankPowerGrid}
\end{figure}

In Figures \ref{fig:SourceRankGrid},\ref{fig:SourceRankPowerGrid} we can see the results of the source location detection experiment for different likelihood estimation functions (AUCDF, AvgTopK and Naive Bayes) on different network structures. The cumulative probability function in these experiments measure the probability of ranking the true source at specific position. 
These results suggest that Naive Bayes and AvgTopK estimators have better performance than the AUCDF estimator. For instance, we can see that in Figure \ref{fig:SourceRankPowerGrid} the Naive Bayes estimator ranks the true source in approximately 80 \% of experiments in top 10 \% of the source list. We have also made a baseline solution which uses random likelihood estimation function to rank the potential sources (see Figures \ref{fig:SourceRankGrid},\ref{fig:SourceRankPowerGrid}). Random likelihood estimation function returns random uniform probability value $[0-1]$ for each node. Note, that the AvgTopK likelihood estimation function tends to give more accurate source localization performance than the Naive Bayes and AUCDF estimation functions. In our experiments we have used the top $k = 5 \%$ of highest scores from pdf in AvgTopK likelihood estimation function.

\begin{figure}[H]
\caption{Cumulative probability distribution of source relative rank for 100 experiments with random initial source on Albert-Barabasi network $(N=5000, M_0 = 5, m = 1)$ for $p=0.6$, $q=0.2$, $T=5$ and different likelihood estimation functions.}
\begin{center}
\includegraphics[scale=0.45]{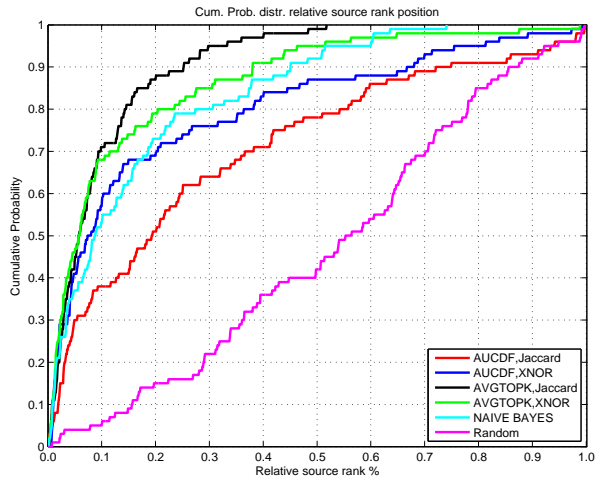} 
\end{center}
\label{fig:SourceRankAB}
\end{figure}

\subsection*{Network structure experiments}
The effects of different network structures on source estimation performance is demonstrated with the following Small-World experiment. We are generating networks from regular lattice ($\beta = 0$) to random networks ($\beta = 1$) with Small-world networks in the middle and observing the performance of source estimators. We measure the area under the cumulative source rank function and observe that the performance of source estimator drops as the average shortest path of network decreases.

\begin{figure}[H]
\caption{ Source location aggregate performance value: area under the cumulative probability of relative source rank (AvgTopk estimator with $\varphi_X()$ similarity function) for 100 experiments on classes of networks (size: $N = 5000$) from regular lattice ($\beta = 0$) to random networks ($\beta = 1$) with Small-world networks in the middle. SIR process has parameters $p = 0.1$, $q = 0.8$ and $T = 7$. Average shortest path is normalized by average shortest path ($\approx 120$) in regular lattice. Average clustering coefficient is normalized by average clustering coefficient ($\approx 0.7$) in regular lattice.}
\begin{center}
\includegraphics[scale=0.45]{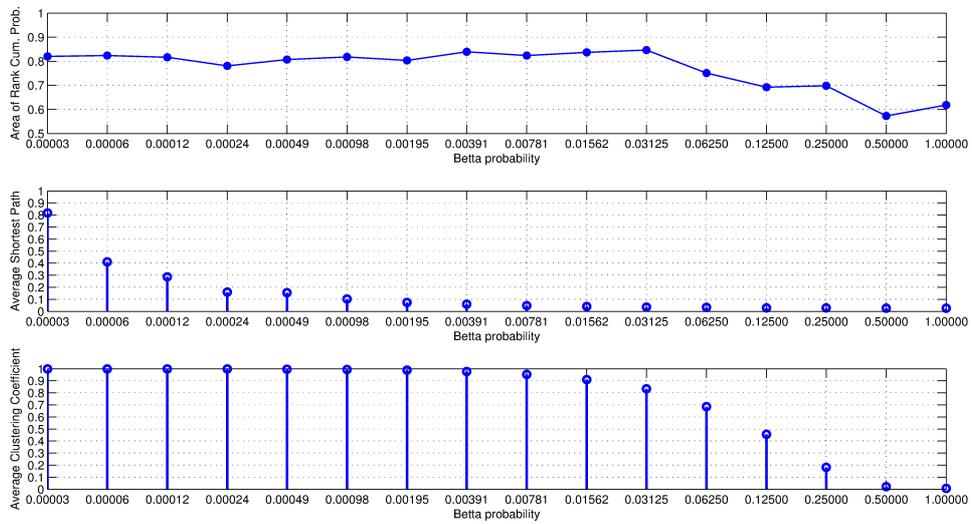} 
\end{center}
\label{fig:SW_case_source_detection}
\end{figure}

\subsection*{Process dynamics experiments}
\begin{figure}[H]
\caption{ Cumulative probability of relative source rank for 100 experiments with random initial source on power-grid network ($N = 4941$) for different parameters $q$. Diffusion like processes are special case of SIR model where recovery parameter $q = 0$ (red line). Experiments were performed with AUCDF likelihood estimation function with $\varphi_X$ similarity function}
\begin{center}
\includegraphics[scale=0.5]{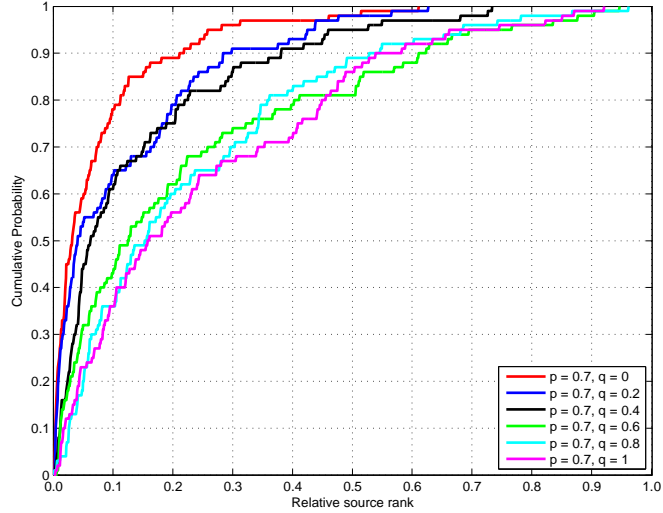} 
\end{center}
\label{fig:SourceRankQCasePower}
\end{figure}

\begin{figure}[H]
\caption{ Cumulative probability of relative source rank for 100 experiments with random initial source on the Albert-Barabasi network $(N=5000, M_0 = 5, m = 1)$ for different parameters $q$. Diffusion like processes are special case of SIR model where recovery parameter $q = 0$ (red line). Experiments were performed with AUCDF likelihood estimation function with $\varphi_J$ similarity function}
\begin{center}
\includegraphics[scale=0.5]{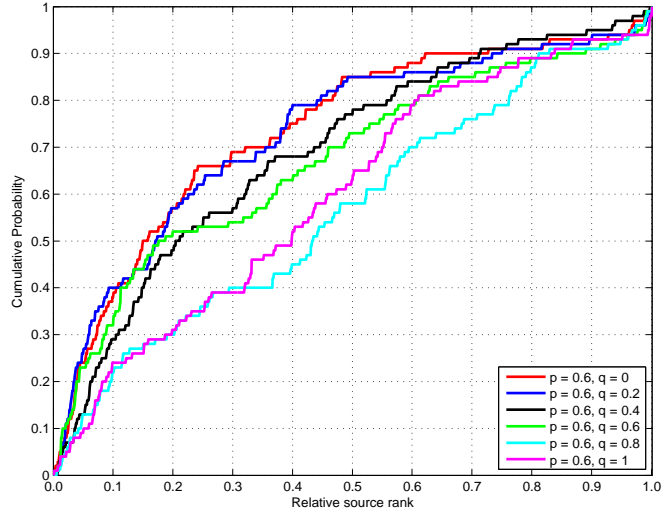} 
\end{center}
\label{fig:SourceRankQCaseAB}
\end{figure}

Finally, we perform a set of experiments to put our source location inference framework into a perspective with recent models for diffusion-like processes published in the literature \cite{Zaman1} \cite{pinto2012}. We illustrate performance of our inference framework on diffusion like processes which can be understood as a limiting case of SIR process in which recovery parameter $q$ is close to or equal zero. 
In Figure \ref{fig:SourceRankQCasePower} and \ref{fig:SourceRankQCaseAB} we can observe that the performance of source estimation algorithms is highest in these conditions. This is expected behaviour which can be interpreted as a consequence of that initial conditions are preserved more in diffusion-like processes. 


\subsection*{Uncertainty experiments}

\begin{figure}[H]
\caption{Cumulative probability of relative source rank for 300 experiments with random initial source on the on power-grid network ($N = 4941$) for $p=0.7$, $q=0.4$, $T = T_0 + \epsilon$, where $\epsilon \sim$ Geometric distribution with different parameters and $T_0 = 15$. }
\begin{center}
\includegraphics[scale=0.45]{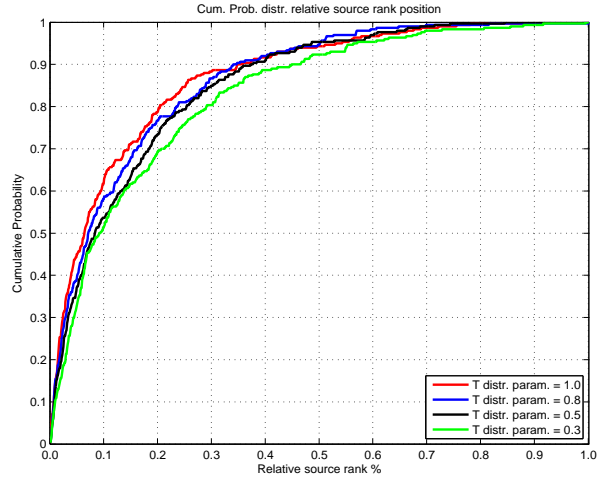} 
\end{center}
\label{fig:TGeometricDistrPower}
\end{figure}

\begin{figure}[H]
\caption{Cumulative probability of relative source rank for 100 experiments with random initial source on the on power-grid network ($N = 4941$) for $p = p_0 + \gamma$, $q = q_0 + \gamma $, $T = T_0 + \epsilon$, where $T_0 = 10$, $p_0 = 0.7$, $q_0 = 0.4$, $\epsilon \sim$ Geometric distribution with parameter $0.5$ and $\gamma \sim$ Normal distribution with parameters $(\mu = 0,\sigma = 0.05)$.}
\begin{center}
\includegraphics[scale=0.45]{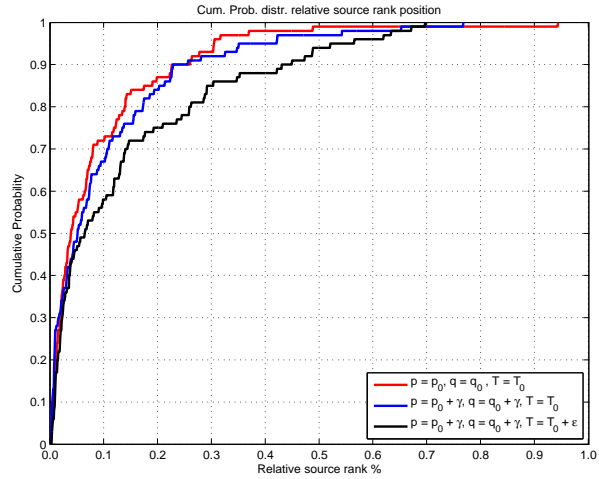} 
\end{center}
\label{fig:PQT_powerGridCase}
\end{figure}

Note that the previous experiments were performed on processes for which the parameters $p$, $q$ and $T$ were degenerative random variable i.e. constants. Now, we demonstrate the effects on performance when the exact values of $p$, $q$ and $T$ are sampled from probability distributions. We model the temporal threshold $T$ as a random variable of the following form: $T = T_0 + \epsilon$, where $\epsilon$ represents the noise from some probability distribution. In Figure \ref{fig:TGeometricDistrPower} we have made a series of experiments where the $\epsilon$ noise was modelled with the Geometric distribution with different parameters. As the variance of noise is increased, the performance of source localization is decreased. We have also made a series of experiments in which the parameters $(p,q)$ were also modelled with the noise: $p = p_0 + \gamma$, $q = q_0 + \gamma $, where $\gamma$ noise was distributed as a Normal distribution with parameters $(\mu,\sigma)$. In Figure \ref{fig:PQT_powerGridCase}, we observe that the performance of source location decreases as the noise of parameters $p$,$q$ and $T$ increases.
This findings suggest that if the predictability is low for parameters $p$, $q$ and $T$ with no noise then predictability can only be lower when the noise is present.
Furthermore, this implies certain limits of predictability for source localization on Small-World networks. \\

\begin{figure}[H]
\caption{Cumulative probability of relative source rank for 100 experiments with random initial source on the on regular grid of size $N = 30x30$ for the ISS spreading process for different parameters $\alpha$, $\beta$, $T=50$ and different fraction of observed nodes (100 \% of realization or 80 \% or 60 \% of random nodes in a realization) in a network.}

\begin{center}
\includegraphics[scale=0.7]{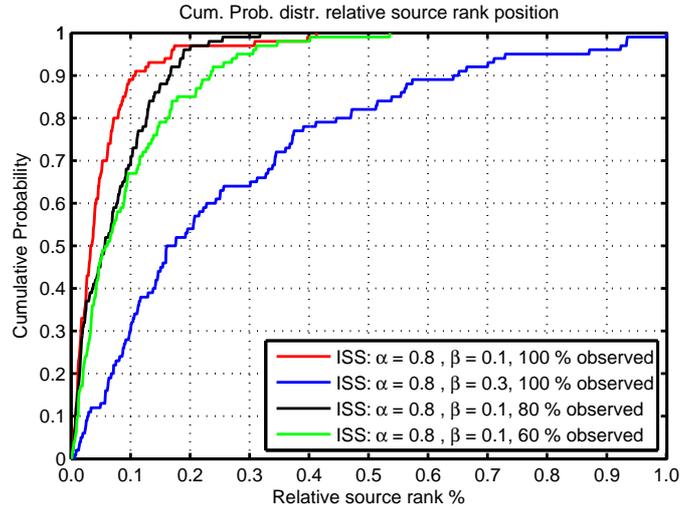} 
\end{center}
\label{fig:ISScase}
\end{figure}

In order to demonstrate the applicability of statistical inference framework for general type of compartment contagion processes, we have made a localization experiments with the information/rumor spreading ISS (ignorant-spreading-stifler) model. The ISS model divide the individuals to three groups: ignorants who have not heard the information/rumor, spreaders who are propagating the information/rumor to ignorants and stiflers who know the information/rumor and are no longer propagating it. The probability of spreading the information/rumor from spreaders to ignorants is $\alpha$ in one discrete time step. If the spreader interacts with other spreader or stifler it turns to stifler state with probability of $\beta$. The infected nodes in the SIR model recovery according to its internal state contrary to the ISS model where spreaders becomes stiflers according to states of its neighbours. In figure \ref{fig:ISScase} we can observe the localization performance of inference framework on ISS model on regular grid for different parameters $(\alpha,\beta)$. Even in case when a fraction of random nodes in a network can be observed the statistical inference framework can localize the initial source (see figure \ref{fig:ISScase}). 

\section{Related work}
\label{relatedWork}
Although the research of epidemic processes on complex networks is very mature the problem of epidemic source detection was formulated very recently. Various  researchers have proposed different solutions to the problem of epidemic source detection which are based on number of assumptions on contact network structures and spreading models. \\
Zaman et. al. formulated a problem, where the rumor spreads with the SI model over network structure for some unknown amount of time and observe information about which nodes got infected. They rise a question who is the most likely source of the rumor and when can they find him. As a solution to the problem of source detection, they developed a rumor centrality measure, which is the maximum likelihood estimator for a regular trees under the SI model. They also obtained various theoretical results about the detection probability on different classes of trees \cite{Zaman1},\cite{Zaman2}. But, when the rumor spreading happens at the general graphs they use the simple heuristics that the rumour spreads along the breadth first search rooted at the source. Dong et. al. also studied the problem of rooting the rumor source with the SI model and demonstrate similar results of asymptotic source detection probability on regular tree-type networks \cite{RegularTreeSI_11}. 
Comin et. al. studied and compared different measures like degree, betweenness, closeness and eigenvector centrality as estimators for source detection \cite{SourceEigenCen_5}. Pinto et. al. also formulated a similar problem of locating the source of diffusion in networks from sparsely places observers \cite{pinto2012}. They also assume that the diffusion tree is a breadth first search, the model of spreading with no recovery and the exact direction and times of infection transfers. Spectral algorithms for detection of initial seed of nodes that best explain given snapshot under the SI model has also been derived \cite{Prakash_8}. \\

Zhu et. al. adopted the SIR model and proposed a sample path counting approach for source detection \cite{Jaccard_7}. They prove that the source node on infinite trees minimizes the maximum distance to the infected nodes. They assume that the infected and susceptible nodes are indistinguishable. Lokhov et. al. use a dynamic message-passing algorithm to estimate the probability that a given node produces an observed snapshot. They use a mean-field-like approximation to compute the marginal probabilities and an assumption of sparse contact network \cite{DMP_0}. \\

Contrary to these approaches, our source estimation approach reduces the assumptions on network structures and spreading process properties.  Our statistical inference framework can also work on arbitrary network structures and with arbitrary compartment spreading processes (SI, SIR, SEIR, ISS, etc.)

\section{Conclusion}
In this paper we have constructed a statistical framework for detecting the source location of an epidemic or rumour spread from a single realization of a stochastic contagion model on an arbitrary network. 
Detecting the source of an epidemic or rumour spreading under a stochastic SIR discrete model, represents an extension of existing research methodologies, mainly focussed on diffusion-like processes. 
Furthermore, this statistical framework can be deployed for different kinds of stochastic compartment processes (ISS, SI, SIR, SEIR) on networks whose dynamical patterns can be described by probability distributions over similarities among realization vectors. 
We have also demonstrated that we can relax even the assumptions on complete knowledge about epidemic realization, contagion process parameters and time with uncertainty.




\section*{Acknowledgements}
The authors would like to thank Professor Yaneer Bar-Yam (New England Complex System Institute, Cambridge, MA 02142, USA) for valuable help in the early stages of the research.

\bibliographystyle{model1b-num-names}
\bibliography{myBib.bib}

\end{document}